\documentclass[conference,a4paper]{IEEEtran}
\IEEEoverridecommandlockouts

\usepackage{enumitem}%

\PassOptionsToPackage{hyphens}{url}\usepackage{hyperref}%
\usepackage[nameinlink, noabbrev, capitalize]{cleveref}%
\hypersetup{
    hidelinks,
    pdftitle={CodeStylist: Supporting Early Undergraduate Programmers with Course-Aware Code Style Feedback},
    pdfauthor={Ethan Dickey, Libra Vento, Peter Kurto, and Andres Bejarano}
}
\usepackage{soul}
\usepackage{graphicx} %
\usepackage{orcidlink} %
\usepackage{xcolor}
\usepackage[numbers,sort&compress]{natbib}
\usepackage{balance}
\def\BibTeX{{\rm B\kern-.05em{\sc i\kern-.025em b}\kern-.08em
    T\kern-.1667em\lower.7ex\hbox{E}\kern-.125emX}}
\begin{document}

\title{CodeStylist: Supporting Early Undergraduate Programmers with Course-Aware Code Style Feedback}

\author{
    \IEEEauthorblockN{%
        Ethan Dickey\orcidlink{0009-0007-3706-5253},
        Libra Vento\orcidlink{0009-0001-6946-3316},
        Peter Kurto\orcidlink{0009-0007-2430-7700}, and
        Andres Bejarano\orcidlink{0000-0003-2611-2855}
    }%
    \IEEEauthorblockA{\textit{Department of Computer Science}\\
        \textit{Purdue University}\\
        West Lafayette, IN 47907, USA}
}

\maketitle
\bstctlcite{BSTcontrol}

\begin{abstract}
    This innovative practice full paper presents CodeStylist, a web application that provides course-standard-aware code style feedback for early undergraduate programming courses. CodeStylist addresses a common instructional gap: students are expected to follow local conventions for naming, formatting, comments, organization, and readability, but feedback on these expectations is often delayed or inconsistent. Unlike generic linters or general-purpose LLM prompts, CodeStylist supports course-specific standards, multi-file submissions, and file- and line-localized explanations intended to guide revision rather than grade correctness. We report a formative expert review with 18 instructional staff from one early undergraduate programming course. Participants explored the prototype using self-selected code artifacts and completed a survey about response quality, anticipated student use, and redesign priorities. Ratings indicated modest perceived utility but limited trust: perceived correctness averaged 60.7\%, response helpfulness averaged 3.50/5, response usefulness averaged 3.33/5, and anticipated student learning averaged 2.61/5. Despite these concerns, 17/18 respondents expected students to use the tool primarily for style checking, often at least weekly.
    Open-ended feedback showed that respondents valued CodeStylist for making implicit course standards more visible, but were concerned about unreliable output, overreliance, and latency or cost. We interpret these findings as evidence that course-aware style feedback is promising as a pre-submission revision aid, but that future versions should combine deterministic rule checks with LLM-generated explanations, rule citations, and stronger verification support.

\end{abstract}


\section{Introduction}\label{sec:intro}
Code style and code quality are central to how programs are read, maintained, and graded, yet in early undergraduate programming courses these expectations are frequently learned implicitly and through delayed, inconsistent human feedback. Prior work shows that automated grading and feedback research has historically emphasized functional correctness, while maintainability- and readability-related feedback remains less common \cite{MesserBKS24}. At the same time, instructors and teaching assistants (TAs) report many ways they would help students improve code quality, but doing so per submission is labor-intensive and may not scale \cite{KeuningHJ19}. Recent large language model (LLM) studies in computing education show promise for generating explanations and formative feedback, while also surfacing reliability, localization, and safety concerns \cite{HellasLLSKKS23,KieslerLK23,AzaizKS24}. More broadly, recent work on scaffolded GenAI adoption in early programming courses suggests that the educational value of such tools depends not only on model capability, but also on whether students are prompted to verify and reflect on suggestions rather than defer to them \cite{dickey2024ailab}. This paper presents \emph{CodeStylist}, a course-aware, LLM-supported code-style feedback tool designed for early programming courses, with an emphasis on course-specific standards, multi-file submissions, and explanation-oriented, localized feedback aligned with instructional staff expectations.

In this paper, we use \emph{code style} to refer to the subset of code quality most visible in early-course standards and rubrics, including naming, formatting, comments, organization, and readability. CodeStylist does not assess functional correctness, algorithmic efficiency, or deeper software design quality. Making this scope explicit is important because correctness-oriented tooling is common in introductory programming, whereas course-specific style expectations often remain implicit and delayed until grading.

\subsection{Motivation and problem statement}
Introductory programming students are asked not only to produce correct outputs, but also to adopt conventions for naming, decomposition, formatting, documentation, and other aspects of code quality. These conventions can be difficult to teach because they blend technical constraints with audience-dependent norms and are not always captured by a single, universally accepted style guide \cite{KirkLT25}. In practice, many courses depend on a combination of rubrics, examples, and TA comments, which can yield delayed feedback and variability across graders \cite{KeuningHJ19}. Such variability matters because educators are not always aligned with each other, and their assessments may diverge from data about novices' actual behaviors and difficulties \cite{BrownA17}. When style is treated as secondary to correctness, students may interpret it as aesthetic or optional rather than as an integral part of communicating intent and supporting maintainability \cite{KirkLT25,MesserBKS24}.

A further structural challenge is scale. Large-enrollment courses often rely on automated assessment for correctness and reserve human effort for higher-level feedback, yet systematic evidence suggests that most automated tools still concentrate on correctness and only a minority address maintainability or readability in depth \cite{MesserBKS24}. Even when educators want to emphasize quality, the time cost of individualized feedback can be prohibitive \cite{KeuningHJ19}. These dynamics motivate tooling that makes expectations explicit, reduces grading variance, and provides timely formative feedback without requiring continuous staff intervention.

\subsection{Pedagogical stakes for novices}
For novices, code quality is not merely professional polish; it can shape what students attend to, what they practice, and what they come to believe ``good code'' means. Empirical work on novices' judgments of readability and structure suggests that novices do not always prefer the same patterns experts do, and that relationships among readability, structure, and comprehension can be complex \cite{WieseRF19}. This complicates purely prescriptive approaches and supports feedback that explains \emph{why} a change improves clarity and how it relates to course learning goals.

Tools that deliver feedback at the moment of writing can also change engagement behaviors. For example, a large-scale deployment of a real-time style feedback tool reported substantially higher student engagement with feedback compared to delayed feedback, and observed that students who viewed feedback often made style-relevant edits \cite{WoodrowMP24}. Importantly, such findings motivate attention to delivery mechanisms and user experience, but they do not by themselves establish learning gains. In this paper, we therefore treat student learning effects as an open empirical question and focus on design, alignment with course practice, and evidence about feedback quality and usability.

\subsection{Design goals}
Code-style tools in education span a range from deterministic linters and static analyzers to data-driven hint systems \cite{LiuP19,HartHMRRT23,MoghadamCYF15,ChoudhuryYF16}. Deterministic checks can be consistent and transparent, but may produce feedback that novices find hard to interpret or prioritize \cite{LiuP19,RechtackovaP26}. Data-driven systems such as AutoStyle and associated hint-generation methods can produce more contextual suggestions, but they still face challenges in generalizing across courses and ensuring that feedback matches a course's specific standards \cite{MoghadamCYF15,ChoudhuryYF16}. Recent education-focused linters aim to make code-quality feedback more adaptable to instructional contexts \cite{RechtackovaP26}.

CodeStylist is designed around four goals that follow from prior work and from the instructional realities of early programming courses. First, it emphasizes
\textbf{course-specificity}: style expectations should be made explicit and configurable to a local coding standard and rubric rather than inferred from generic conventions \cite{KirkLT25,KeuningHJ19}. Second, it emphasizes
\textbf{project realism}: the tool should support multi-file submissions and course-relevant program structure, since novice assignments often extend beyond single-script exercises \cite{MesserBKS24}. Third, it emphasizes
\textbf{explanatory feedback}: students benefit not only from being told that something may be wrong, but also from seeing why it matters and how it relates to the course's expectations \cite{AzaizKS24,KeuningHJ19}. Fourth, it emphasizes
\textbf{localization and usability}: feedback should be anchored to specific files, lines, or code regions so that students can inspect and revise their work with lower cognitive overhead \cite{AzaizKS24,VasconcelosBFLV25}.

Taken together, these goals position CodeStylist as a course-aware, revision-oriented feedback tool for early undergraduate programming. Its purpose is to make local style expectations more visible while students are still revising, to support realistic submission formats, and to provide feedback that is both interpretable and actionable. In this way, CodeStylist is intended to help students connect written coding standards to concrete decisions in their own code rather than encounter style feedback only after grading or in one-off staff interactions.

\subsection{Scope of evidence and claims}
LLM-based feedback can be helpful but unreliable. Studies of LLM responses to beginners' help requests show frequent partial correctness, missed issues, and false positives, as well as sensitivity to prompt characteristics such as language choice \cite{HellasLLSKKS23}. Studies focused on generating feedback for programming submissions similarly report strengths in structure and detail alongside risks such as contradictory statements and localization errors \cite{AzaizKS24}. Work proposing hybrid pipelines highlights one route to mitigate these risks by combining stronger ``tutor'' models with validation steps that simulate student utility \cite{PhungPS0CGSS24}.

Accordingly, this paper presents CodeStylist as an innovative practice for translating a local course coding standard into pre-submission feedback for novice programmers. The paper contributes a design account of a course-aware workflow for localized style feedback and formative expert review evidence about where the current prototype appears instructionally useful, where trust breaks down, and what redesign priorities follow. Because the evidence comes from instructional staff review rather than student outcome data, we frame these contributions as evidence about alignment, perceived usefulness, and instructional credibility rather than as evidence of student learning gains.

\section{Related Work}\label{sec:related_work}
\subsection{Automated style and code-quality feedback systems for novices}
Early systems explored generating formative style feedback at scale by leveraging patterns in student solutions and instructor guidance. AutoStyle targeted holistic style feedback and demonstrated how data-driven approaches can move beyond correctness-only critiques \cite{MoghadamCYF15}. Related work in intelligent tutoring explored scale-driven hint generation for coding style, framing style feedback as a hint-generation problem and operationalizing instructor-authored guidance \cite{ChoudhuryYF16}. Subsequent work also investigated pedagogy around recognizing and implementing good style, underscoring that style instruction requires more than a checklist and benefits from feedback that supports interpretation and revision \cite{WieseYCSF17}.

More recently, education-oriented tooling has emphasized practical deployment in courses. Eastwood-Tidy provided C linting for style assessment in programming courses \cite{HartHMRRT23}, while Learning with Style investigated how to better structure automated style feedback to improve student responsiveness \cite{SalibaSOCQ24}. EduLint similarly positions itself as a versatile code-quality feedback tool, highlighting configurability and educational usability concerns \cite{RechtackovaP26}. In Python contexts, integrating static analysis into CS1 has been studied both for student perceptions and for how such tools fit within course workflows \cite{LiuP19}. Across these systems, common tradeoffs include the completeness of rule coverage, student comprehension of messages, and the extent to which feedback fits a particular course's standards and assignment structures \cite{KeuningHJ19,KirkLT25}.

\subsection{LLM-based programming feedback and hybrid deterministic plus generative designs}
A growing body of work examines LLMs for formative programming feedback. In one study, researchers explored the potential for LLMs to generate formative programming feedback and highlighted both opportunities and caveats for using such feedback in educational settings \cite{KieslerLK23}. Other work analyzing LLM responses to authentic beginner help requests documented that LLMs often identify some issues but frequently miss others and may produce false positives, with only modest performance degradation for non-English prompts \cite{HellasLLSKKS23}. A later work studying GPT-4-based feedback generation for submissions similarly reported improvements in structure and consistency relative to earlier models, alongside issues such as contradictory feedback and imperfect fault localization \cite{AzaizKS24}.

In parallel, large-scale deployments have reported how LLM-based style feedback can change interaction patterns with feedback: a recent study described a real-time style feedback tool deployed in a large global CS1 course and reported increased feedback engagement and subsequent style-related edits, while also discussing safety and bias mitigation techniques needed for such interventions \cite{WoodrowMP24}. Prior work has also proposed hybrid architectures that use stronger models for hint generation and weaker models for validation, explicitly treating validity checking as a design requirement rather than an afterthought \cite{PhungPS0CGSS24}. These results collectively motivate CodeStylist's emphasis on course-aware, explanation-oriented feedback while underscoring the need for stronger validation, clearer localization, and more trustworthy pipelines \cite{AzaizKS24,PhungPS0CGSS24}.

\subsection{Interface, localization, and trust in feedback tools}
Even accurate feedback can be ineffective if it is poorly presented. Prior work in CS education emphasizes that automated feedback must be actionable and aligned with novice needs, including clarity about what to change and where \cite{KeuningHJ19,LiuP19}. LLM-based feedback magnifies this requirement because generative text can be fluent yet wrong, and because students may over-trust authoritative explanations \cite{HellasLLSKKS23,AzaizKS24}. Recent HCI research on uncertainty highlighting in AI code completions suggests that interface-level signals can help users calibrate trust and attend to potentially unreliable regions of AI output \cite{VasconcelosBFLV25}. For educational feedback contexts, these insights motivate presenting localized, inspectable feedback rather than undifferentiated prose, and motivate designs where deterministic checks anchor the feedback to concrete artifacts (lines, files, rule IDs) \cite{VasconcelosBFLV25,RechtackovaP26}.

\subsection{Synthesis of themes motivating CodeStylist}
Across this literature, four themes are especially relevant to the present work. First, code style expectations are often implicit, variable, and locally defined, which makes course-aware representations of style particularly important in educational settings \cite{KeuningHJ19,KirkLT25}. Second, the timing of feedback matters: students are more likely to engage with style feedback when it is available while they are still revising rather than after grading \cite{MesserBKS24,WoodrowMP24,SalibaSOCQ24}. Third, different feedback technologies offer different strengths. Static analysis and educational linters provide consistency for objective checks, while generative approaches can offer more flexible explanations but introduce new reliability risks \cite{LiuP19,HartHMRRT23,RechtackovaP26,HellasLLSKKS23,AzaizKS24}. Fourth, both the education and HCI literature suggest that trust depends on inspection and localization: students need to see where a problem occurs, what rule it relates to, and whether the feedback is trustworthy enough to act on \cite{AzaizKS24,VasconcelosBFLV25}.

CodeStylist is motivated by the intersection of these themes.
Its novelty is not that course specificity, project realism, explanation, or localization is individually unprecedented: educational linters and static analyzers address consistency and localization, style-feedback systems address revision-oriented hints, and LLM-feedback studies address explanation, but these goals are rarely combined around instructor-provided local standards and multi-file pre-submission use.
Rather, CodeStylist combines them in a course-aware pre-submission workflow: local standards are treated as explicit configuration artifacts, multi-file submissions are supported, suggestions are returned in a structured file/line format, and the LLM is used to explain possible violations of the selected course standard rather than to provide generic programming advice. This framing leads to the central question of the present paper: whether instructional staff view the current prototype as useful, trustworthy, and instructionally aligned enough to justify further development toward classroom deployment.

\section{CodeStylist: The Practice and Context} \label{sec:codestlyist}
CodeStylist is a standalone prototype intended for optional pre-submission use in early undergraduate programming courses that employ explicit local coding standards. Students upload one or more source files, select the relevant course standard and programming language, and receive structured, line-referenced style suggestions organized by file. The instructional purpose of the tool is to help students inspect code against course expectations while they are still revising, not to judge functional correctness or replace staff feedback.

This context matters because many early undergraduate programming courses expect students to follow local standards for naming, comments, organization, and readability, yet detailed feedback on those expectations is often delayed until grading or office hours. CodeStylist was therefore conceived as a way to move some of that feedback earlier in the workflow by translating the existing course standard into more actionable pre-submission guidance.

\subsection{Instructional goals}
The design of CodeStylist followed four instructional goals. First, the tool should make implicit or easy-to-overlook style expectations more explicit for novice programmers. Second, it should be aligned with local course standards rather than with a generic notion of good style, since early courses often differ in what they emphasize and how they operationalize those expectations. Third, it should support the kinds of submissions common in foundational programming courses, including multi-file work. Fourth, it should provide guidance that is specific enough to support revision, while still remaining a support tool rather than an automated grader or correctness checker.

These goals shaped both the interface and the feedback logic. CodeStylist was built to focus on style-related issues such as naming, comments, formatting, organization, and adherence to the course standard. It was not intended to judge functional correctness or to generate full solutions. That distinction matters pedagogically. The instructional role of the tool is to help students inspect and revise code with the course standard in mind, not to substitute for problem solving or staff judgment.

\subsection{Workflow and feedback generation}

In the evaluated prototype, a user uploads one or more source files, selects the relevant programming language and course coding standard, and submits the bundle for analysis. The system numbers the lines within each file and constructs a model request containing the line-numbered code, selected language, and selected course standard. In this version, ``LLM-supported'' means that these materials were sent to GPT-4o with instructions to identify possible violations of the selected standard and return structured feedback containing the relevant file, line range, issue, suggestion, and explanation. A course configuration consisted of the programming language and an instructor-provided coding standard manually entered from course documents; the prompt template and configuration format used in the evaluation are available in the project repository.\footnote{\href{https://github.com/davento/codeStylistTA}{CodeStylist project repository}.} The evaluated version did not include deterministic validation of objective rules, a limitation that likely contributed to the reliability concerns reported below.

\begin{figure*}%
    \centering
    \includegraphics[width=\linewidth,keepaspectratio]{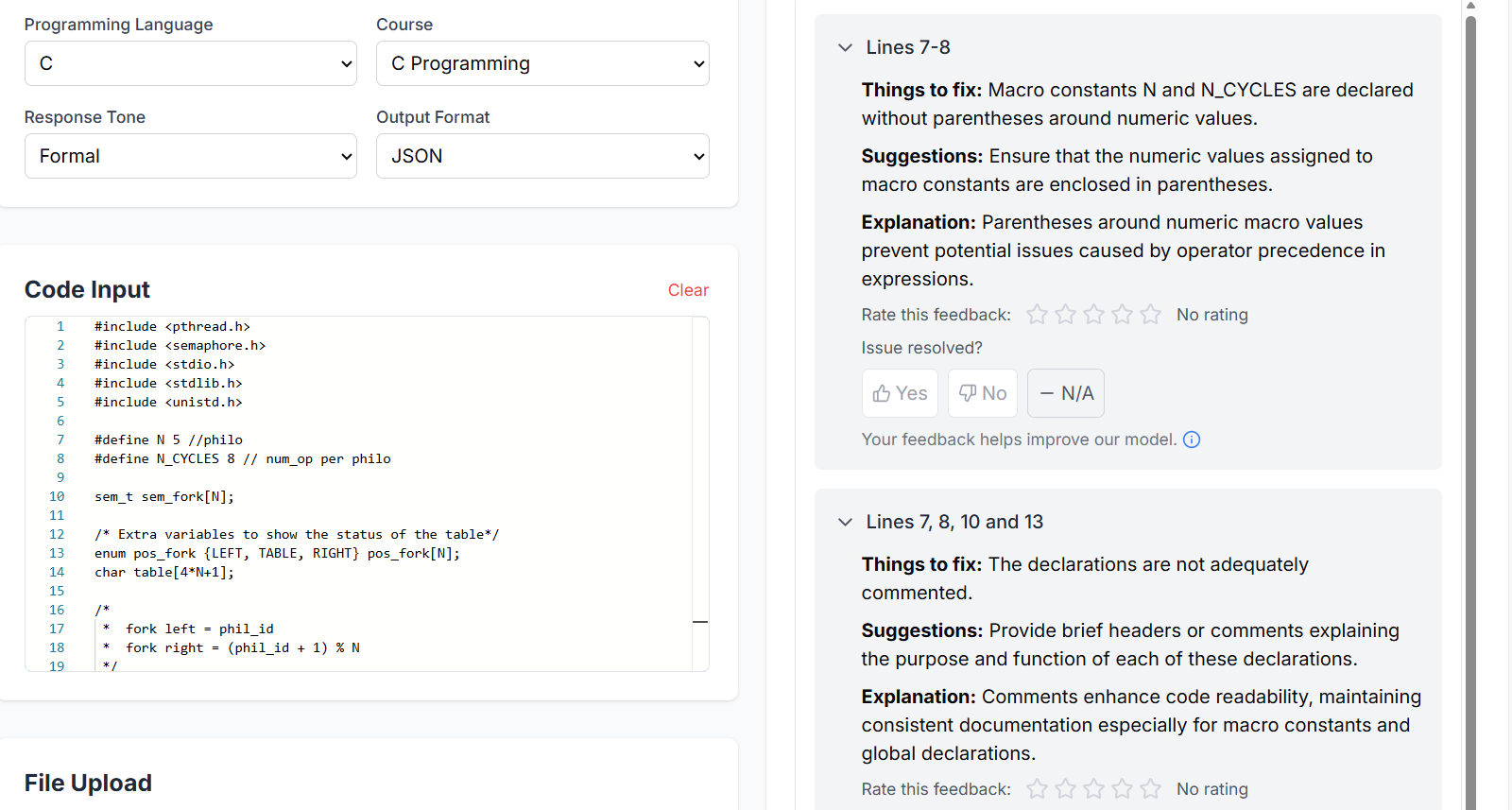}
    \caption{CodeStylist interface during formative review. Users select a course-specific coding standard and programming language, inspect source code with line numbers, and receive file- and line-localized style suggestions with brief explanations to support pre-submission revision.}
    \label{fig:stylist_workflow}
\end{figure*}

The interface renders the returned suggestions by file and line so that users can inspect each issue in context, compare it against the local standard, and decide whether revision is warranted (\Cref{fig:stylist_workflow}). This workflow reflects two central design commitments. First, feedback should be course-aware rather than based only on generic style conventions. Second, feedback should be localized and inspectable rather than presented as undifferentiated prose about the submission as a whole. The design goal is therefore not to replace linters, grading, or staff judgment, but to complement existing course infrastructure with pre-submission, explanation-oriented feedback tied to local expectations.

\subsection{Intended instructional scope}

CodeStylist is intended for early undergraduate programming courses in which students are expected to follow an explicit local coding standard and where style is visible enough to matter for feedback, revision, or assessment. Rather than being tied to a single course, the tool is designed to be configured to course-specific standards, languages, and submission structures across CS1- and CS2-level contexts. During development, the prototype was configured against multiple early-course standards, although the formative expert review reported here focuses on one evaluated configuration. In these settings, students often receive delayed or inconsistent guidance about how a written standard applies to their own submissions. CodeStylist was therefore conceived as an optional pre-submission aid that can translate local standards into more actionable, submission-specific style feedback while students are still revising.

At the time of this study, CodeStylist was a standalone prototype rather than a required course component. It was not integrated into grading, and it was not positioned as an authoritative checker. The present evaluation examined one course-specific configuration of the prototype with instructional staff from a single early undergraduate course in order to assess whether the practice appeared instructionally useful and how it should be refined before broader student use. This framing is important for interpreting the results reported later in the paper. The study does not test whether CodeStylist improves student learning outcomes. It examines whether instructional staff view the current prototype as potentially useful, how they expect students to use it, and what design changes are needed before classroom deployment at larger scale.

\section{Formative Expert Review Method} \label{sec:method}
Before student deployment, we conducted a formative expert review of CodeStylist. We chose instructional staff as reviewers because they work at the intersection of student need and course policy. They routinely interpret the course standard, read novice code, explain style expectations, and see where students struggle during revision. This makes them well positioned to judge both whether the tool's feedback aligns with instructional goals and whether students are likely to use it productively.

The review was designed to address three practical questions: (1) How do instructional staff judge the quality of the prototype's feedback? (2) How do they expect students to use the tool under current course policies? and (3) What benefits, risks, and redesign priorities do they identify before student-facing deployment?

\subsection{Participants and setting}
Participants included 18 instructional staff from Purdue University's Fall 2025 offering of CS240, an early undergraduate C programming course: 9 undergraduate TAs, 8 graduate TAs, and 1 instructor.
They had supported the course for one to five semesters. All participants therefore brought prior experience with the course's local coding standard and common student difficulties around style, revision, and code comprehension.

This participant group was purposively appropriate for a formative review. Undergraduate and graduate teaching assistants regularly work with student submissions and student questions, while the instructor provides a course-level perspective on standards and implementation. Taken together, these roles offered an expert view of both the student-facing and instructional-facing implications of the prototype.

\subsection{Procedure}
We invited instructional staff associated with the course used for the present evaluation to interact with the current CodeStylist prototype as configured for that course.
Participants took part in a 40-minute formative review session. At the start of the session, they received a brief introduction to CodeStylist, including its intended purpose and the perspective from which they were being asked to evaluate it. Participants were then invited to explore the prototype independently.

To test the tool, participants uploaded code documents of their own choosing. These included de-identified student documents available to them through prior instructional experience as well as self-generated examples created for testing. Participants interacted with the prototype, reviewed the resulting feedback, and completed the survey at the end of the session.

This procedure prioritized ecological validity over standardized benchmarking. We wanted reviewers to stress-test the prototype with artifacts they considered representative of real course use, rather than constrain them to a fixed evaluation set. The tradeoff is that the resulting ratings reflect authentic exploratory use rather than controlled performance comparison across identical submissions. We therefore interpret these results as formative judgments about instructional fit, trust, and likely classroom role, not as a benchmark of model accuracy.

\subsection{Measures}
The survey began with two background questions: participant role and number of semesters supporting the course. It then asked a set of tool-focused questions combining ordinal ratings, structured response items, and open-ended prompts.

The closed-ended items captured several distinct constructs. Participants estimated the correctness of CodeStylist's feedback as a percentage; this item asked for participants' subjective judgment of how much of the feedback they viewed as aligned with the course standard during exploratory use, not benchmark accuracy against a common labeled dataset. Participants also rated how helpful and useful the tool's answers were for their own review of the prototype. To understand anticipated classroom use, participants indicated how they thought students would use the tool and how frequently they expected students to use it under current course policies. Finally, they rated how helpful the tool would be for students, how useful it would be in solving students' problems, and how much students were likely to learn from it if used for its intended purpose.

The open-ended items were designed to complement those ratings with richer instructional judgment. Participants were asked about potential benefits, potential challenges, whether the tool might be useful in other courses, interface and tool feedback, and any additional comments they wished to share. Together, these questions allowed the study to capture both immediate reactions to prototype quality and broader judgments about classroom fit.

\subsection{Analysis}
We analyzed the closed-ended responses descriptively. For perceived correctness, we report summary statistics such as averages, medians, and ranges. For the ordinal rating items, we report means and medians and interpret responses relative to the neutral midpoint in the instrument. We also summarize the distribution of positive, neutral, and negative responses to avoid relying on means alone. Responses about anticipated student use were summarized with counts and percentages.

Because the sample was small and the purpose of the review was formative, we did not conduct inferential statistical tests. GTA and UTA results are reported descriptively to highlight possible role-based patterns, not to support broader population claims. The instructor response is shown for transparency but is not interpreted as a subgroup on its own.

We analyzed the open-ended responses using a lightweight descriptive coding process oriented toward redesign. The authors reviewed all comments, grouped them into categories relevant to practice improvement, and used counts only to indicate the prominence of concerns when helpful (e.g. perceived benefits, perceived risks, transfer conditions, and interface needs). Because the responses were brief and the purpose of the study was formative rather than theory building, we treat these categories as design-oriented summaries of expert concerns rather than as fully developed qualitative themes.

\section{Results}\label{sec:results}
\Cref{tab:expert-review-quant} summarizes the six quantitative items from the expert review. Because the five-point items were ordinal and the sample was small, we report medians and category counts alongside means. For these items, 3 represented a neutral midpoint, 4--5 positive responses, and 1--2 negative responses. The respondent pool included 9 undergraduate TAs (UTAs), 8 graduate TAs (GTAs), and 1 instructor. Respondents had supported the course for 1--5 semesters; 11 reported 1 semester of experience, 1 reported 2 semesters, 2 reported 3 semesters, 2 reported 4 semesters, and 2 reported 5 semesters.

\begin{table*}[t]
    \caption{Descriptive summary of quantitative responses. GTA and UTA columns report subgroup means only. Given the small, purposive sample, subgroup comparisons are descriptive rather than inferential.}
    \label{tab:expert-review-quant}
    \centering
    \small
    \begin{tabular}{lccccl}%
        \hline
        Measure & Overall & GTA & UTA & Instr. & Overall detail \\
        \hline
        Perceived correctness (\%) & 60.7 & 71.5 & 53.9 & 35.0 & Median 64.5; range 4--100 \\
        Response helpfulness & 3.50 & 3.75 & 3.33 & 3.00 & Median 4; positive/neutral/negative = 12/3/3 \\
        Response usefulness & 3.33 & 3.63 & 3.11 & 3.00 & Median 4; positive/neutral/negative = 11/2/5 \\
        Helpfulness for students & 3.56 & 3.75 & 3.33 & 4.00 & Median 4; positive/neutral/negative = 13/1/4 \\
        Usefulness for students & 3.33 & 3.13 & 3.44 & 4.00 & Median 4; positive/neutral/negative = 11/4/3 \\
        Learning from the tool & 2.61 & 2.63 & 2.56 & 3.00 & Median 3; positive/neutral/negative = 3/8/7 \\
        \hline
    \end{tabular}
\end{table*}

\subsection{Perceived response quality and correctness}
Perceived response quality was mixed. Perceived correctness estimates averaged 60.7\% overall, with a median of 64.5\% and a wide range from 4\% to 100\%. For the two items that asked respondents to evaluate the quality of the tool's answers directly, helpfulness was slightly above neutral ($M = 3.50$, $Mdn = 4$), while usefulness was near-neutral ($M = 3.33$, $Mdn = 4$). The distribution of responses is important here: although the medians for both items were 4, five respondents rated response usefulness negatively and three rated response helpfulness negatively. Thus, many respondents saw practical value in the feedback, but a substantial minority did not find it sufficiently reliable or useful.

A descriptive role-based split was also apparent. GTAs rated response quality more favorably than UTAs on all three measures shown in \Cref{tab:expert-review-quant}: perceived correctness (71.5 vs. 53.9), helpfulness (3.75 vs. 3.33), and usefulness (3.63 vs. 3.11). In contrast, UTAs rated usefulness for students more favorably than GTAs (3.44 vs. 3.13). Because subgroup sizes were small, participants used self-selected artifacts, and the study was not designed for between-group comparison, we interpret these differences only as possible redesign signals rather than as evidence of role-based effects. They do, however, identify a meaningful pattern in who appears more skeptical of the current prototype. One UTA response was especially negative, rating correctness at 4\%, response usefulness at 2, anticipated learning at 1, and predicted student frequency of use as ``Never.'' This case contributed to the wide spread in the UTA responses and underscores the heterogeneity of opinion about the tool's current quality.

\subsection{Anticipated student use and value}
Respondents generally believed that students would use CodeStylist for its intended purpose. Seventeen of the 18 respondents anticipated code style checking as a student use case; 12 selected code style checking only and 5 selected code style checking together with syntax debugging. One respondent selected ``Other'' and wrote that students would not use the tool because the linter already sufficed for the task. When asked about likely frequency of use under current course policies, 14 of 18 respondents (77.8\%) predicted at least weekly use: 8 selected ``Once a week,'' 5 selected ``2--3 times/week,'' and 1 selected ``Every day.'' The remaining responses were 2 ``Only once or twice'' and 2 ``Never.'' UTAs most often predicted once-weekly use (5/9), whereas GTAs were split between once-weekly and 2--3 times/week predictions (3 each).

Expectations for student-facing value were somewhat more positive than evaluations of the tool's raw response quality, but they were still modest. Perceived helpfulness for students was slightly above neutral ($M$ = 3.56, $Mdn$ = 4), with 13 positive responses and 4 negative responses. Perceived usefulness for students was near-neutral to slightly positive ($M$ = 3.33, $Mdn$ = 4), with 11 positive, 4 neutral, and 3 negative responses. This item also reversed the earlier subgroup pattern: UTAs rated usefulness slightly higher than GTAs did (3.44 vs.\ 3.13), despite rating the tool's response quality less favorably.

Anticipated student learning was the weakest quantitative outcome. Learning from the tool averaged 2.61, below the neutral midpoint, with a median of 3. Only 3 respondents rated the learning impact positively, while 7 rated it negatively and 8 chose the neutral option. Taken together, these results suggest that respondents saw CodeStylist as more promising for helping students catch style issues than for reliably teaching the underlying principles behind those issues.

\subsection{Open-ended responses}
We grouped recurring ideas across the open-ended responses as descriptive themes. A single response could receive multiple codes.

For perceived benefits, the dominant theme was that CodeStylist could make code style standards more visible and actionable. Respondents described the tool as helping students identify likely standard violations, understand what the code standard is asking for, and revise before submission so they do not lose points on style-related rubric criteria. A second recurring benefit was explanatory value: several respondents described the tool as offering a ``second pair of eyes'' or a more interpretable explanation than a linter. A smaller set of responses emphasized speed and scalability, such as providing feedback faster than TAs can in large courses.

For perceived challenges, reliability clearly dominated. In our descriptive coding of Q14, 8 of the 14 non-empty responses explicitly mentioned incorrect, inconsistent, hallucinated, or incomplete feedback. Four mentioned overreliance or reduced learning, and 3 mentioned latency or token cost. One UTA wrote that the tool was ``not accurate towards the actual code standard it was tested against'' and worried that this could negatively affect students if they trusted incorrect guidance. Another respondent warned that unreliable output could create additional TA work when students bring back tool-generated feedback that conflicts with course expectations.

The transferability and interface responses added useful boundary conditions. Respondents saw the strongest fit in courses with explicit coding standards, especially early courses and courses in which students are learning a new language, and in settings where there is no strong existing style-checking infrastructure. At the same time, some respondents were skeptical about value in contexts where style is not visibly graded or where IDE and linter support already cover much of the need. Interface comments were strikingly consistent. Five responses explicitly requested tighter linkage between feedback and code, such as a split view, clickable line numbers, or highlighted offending lines. Additional requests included issue categorization, citation of the specific rule in the coding standard, and avoiding prescriptive fixes when the pedagogical goal is rule learning rather than answer substitution. Respondents also reported robustness problems, including incorrect line references, poor handling of comments in larger files, and submission glitches.

\section{Discussion} \label{sec:discussion}
\subsection{Task utility and learning value diverged}
The clearest pattern in the expert review is a separation between immediate task utility and anticipated learning value. Most respondents expected students to use the tool regularly for style checking, and the medians for both student helpfulness and student usefulness were 4. At the same time, perceived correctness averaged only 60.7\%, and anticipated learning was the only five-point outcome with a below-neutral mean. In practical terms, respondents saw more promise in CodeStylist as a revision aid than as a learning intervention in its current form.

That pattern is consistent with prior work on automated and AI-supported programming feedback. Earlier research on automated style feedback showed that timely, actionable style support can help students revise toward better code style and recognize style issues more effectively \cite{MoghadamCYF15,WieseYCSF17}. More recent work has shown that LLM-based style feedback can also support timely, scalable feedback workflows \cite{WoodrowMP24}. At the same time, recent studies consistently caution that LLM-generated programming feedback can still be misleading, contradictory, or insufficiently calibrated for novice learners, which makes validation and careful interface design essential before classroom deployment \cite{KieslerLK23,AzaizKS24,PhungPS0CGSS24}. Our results place CodeStylist squarely in that space: the need is real, the use case is recognizable, but trust remains the main obstacle.

\subsection{Reliability is the central adoption barrier}
The quantitative and qualitative results converged on the same issue. Perceived correctness estimates were highly variable, and reliability was by far the most common challenge theme in the open-ended responses. This matters pedagogically as well as technically. A style tool that produces plausible but wrong feedback can waste student time, create confusion about course standards, and increase TA workload if students ask staff to arbitrate between the tool and the official standard.
The UTA who described the tool as ``not accurate towards the actual code standard it was tested against'' illustrates the most important pedagogical risk: students may have difficulty recognizing when confident feedback conflicts with course expectations.

For this paper's claims, this means CodeStylist should not be framed as an authoritative grader or tutor. A more accurate framing is that the current prototype is a course-aware suggestion generator that can surface possible style issues, but whose output still requires verification. That is still a meaningful instructional contribution, especially in contexts where style expectations are important but often implicit. It is simply a narrower and more defensible claim.

\subsection{What the GTA and UTA split adds}
The GTA and UTA split is useful as a design-relevant pattern, but it should be interpreted cautiously. GTAs were consistently more favorable on response quality, while UTAs were slightly more favorable on usefulness for students. With a sample this small, several explanations remain plausible: differences in expectations, differences in chosen test cases, differences in day-to-day proximity to student workflow, or simple response heterogeneity. We also cannot cleanly separate role from other local factors. For example, GTAs in this sample had more semesters of course experience on average than UTAs did, but the data do not support any stronger causal reading than that.

Even so, the split is informative. One plausible interpretation is that GTAs were more willing to tolerate imperfect output if the tool helps surface standards earlier and reduces repetitive feedback work, whereas UTAs were more sensitive to the immediate friction students would feel when feedback is wrong or redundant with existing tools. That interpretation fits the open-ended responses, especially the UTA comments about linters, IDE support, and objective rule checks. For the design of CodeStylist, the implication is simple: the tool needs to satisfy both pedagogical and workflow-centered expectations if it is to be adopted.

\subsection{Design implications}
The first design implication is to separate objective detection from explanation. The most common challenge theme was unreliable output, and several respondents explicitly noted that some objective checks should not be left to an expensive and error-prone model when they can be implemented directly in code. For CodeStylist, that suggests a hybrid architecture: deterministic checks for objective rules, with the LLM reserved for explanation, contextualization, and prioritization. This recommendation also aligns with recent work showing that validation layers can improve the reliability of generated programming feedback \cite{PhungPS0CGSS24}.

The second design implication is to make verification easier than compliance. Respondents repeatedly requested split-view code display, exact line highlighting, rule citation, and clearer issue categories. These requests are not merely interface polish. They directly support student judgment. If students can see more clearly where a problem is located and which rule it is supposed to violate, they are in a much better position to decide whether the tool is correct. This matters because recent work on LLM-generated programming feedback shows that outputs can be more structured and polished than earlier generations of tools while still containing contradictions or localization errors \cite{AzaizKS24}.

The third design implication is to keep the tool tightly scoped to style coaching. Some respondents anticipated syntax-debugging use, but the current evidence supports CodeStylist most strongly as a course-aware style assistant. Expanding the scope before reliability improves would likely increase misuse and disappointment. A narrower, rule-grounded interaction is more consistent with the instructional problem CodeStylist is trying to solve, namely making local, instructor-defined style expectations visible early enough for students to act on them.

Finally, if the pedagogical goal is learning rather than mere correction, the tool should shift from suggestion delivery toward guided interpretation. Rather than providing an over-prescriptive fix, especially when confidence is low, future versions should cite the relevant section of the coding standard, explain why the issue may apply, and prompt students to decide how to revise. This recommendation is especially well aligned with the below-neutral learning ratings in this study. If the tool does too much of the student's interpretive work, it may help with compliance while doing little to build transferable judgment.

\subsection{Threats to validity, scope conditions, and next steps}
The open-ended responses also point to clear scope conditions. Respondents saw the strongest fit in early programming courses, in courses where students are learning a new language, and in contexts where explicit style standards matter for assessment. They were less optimistic where style is not visibly graded or where linters and IDE support already cover most of the need. CodeStylist is therefore not best framed as a universal AI assistant for programming education, but rather as an infrastructure for turning local, instructor-authored style expectations into earlier and more interpretable feedback.

This study has important limitations. It is a formative expert review from a single course context with 18 respondents. The quantitative findings are descriptive, the open-ended responses are short, and the subgroup differences are not inferential. We also do not yet have student log data, revision data, or blinded evaluations of changes in code quality. The next evaluation should therefore move from perception to behavior: which suggestions students act on, whether rule-cited and line-localized feedback is trusted differently than the current prototype, and whether revisions made with the tool improve rubric-based style scores without encouraging overreliance. A particularly useful next comparison would test the current LLM-centered prototype against a hybrid version that pairs deterministic rule detection with LLM explanations.

In addition, participants evaluated the prototype using self-selected code artifacts rather than a common standardized set of submissions. Because artifact type and source were not systematically coded for comparison, we cannot determine whether ratings varied by artifact source, size, completeness, or difficulty. The quantitative ratings should therefore be interpreted as judgments from authentic exploratory use rather than as controlled performance comparisons.

\section{Conclusion}\label{sec:conclusion}
CodeStylist was motivated by a familiar problem in early undergraduate programming: code style matters for readability, maintainability, and assessment, but students often receive feedback on style late, inconsistently, or in ways that are difficult to act on. This paper presented CodeStylist, a course-aware tool for making local style expectations more visible before submission, and reported a formative expert review with instructional staff who regularly support novice programmers.

The results suggest a clear but bounded contribution. Instructional staff saw meaningful promise in CodeStylist as a style-checking and revision aid, especially for helping students identify likely violations of a course standard and for making implicit expectations more explicit. At the same time, they expressed limited trust in the current prototype's reliability and did not view it as a strong learning intervention in its present form. In other words, the expert review indicates that the instructional need is real and that the general direction is worthwhile, but that adoption depends on improving trust, localization, and verification support.

Taken together, the findings point to several practical implications for AI-supported style feedback in computing education. Course-aware feedback appears most promising when it is grounded in explicit local standards, localized to specific parts of the submission, and framed as support for inspection and revision rather than as authoritative judgment. The review also suggests that future versions should better separate objective detection from explanation, cite relevant rules from the coding standard, and make it easier for students to verify whether a suggestion is correct.

This study is necessarily limited to a formative expert review in a single course context. The next step is a student-facing evaluation that examines actual use, revision behavior, and changes in code quality under blinded assessment. Even so, the present paper contributes an empirically grounded design account of what it takes to make course-aware code-style feedback more instructionally credible for early undergraduate programming.

\section*{Acknowledgments}
This work was funded by Purdue's Innovation Hub (IH-AI-23002) and the Department of Computer Science through the GoBoiler program. OpenAI's ChatGPT (GPT-5.5) was used for sentence-level language editing in the abstract and the body; all content was reviewed and approved by the authors.

\balance
\bibliographystyle{IEEEtran}
\bibliography{refs}

\end{document}